\documentstyle[prd,tighten,aps]{revtex}

\newcommand{\rf}[1]{(\ref{eq:#1})}

\begin{document}
\draft

\twocolumn[\hsize\textwidth\columnwidth\hsize\csname
@twocolumnfalse\endcsname
\title{Geometry of Brane-Worlds}
\author{M. D. Maia}
\address{Universidade de Brasilia, Instituto  de Fisica, 
 Brasilia. D.F. 7919-970,\\maia@fis.unb.br} 
\author{ Edmundo M. Monte }
\address{Universidade Federal da Paraiba,
Departamento de Fisica,  Jo\~ao Pessoa, Pb. 58059-970\\edmundo@fisica.ufpb.br}

\date{\today}

\maketitle

\begin{abstract}
The most general  geometrical scenario in which   the brane-world program  can be 
implemented is  investigated.  The basic requirement is that it  should be  consistent with the   confinement of gauge interaction, the existence of quantum states  and the  embedding in  a   bulk  with  arbitrary  dimensions, signature and  topology.   It is found that  the  embedding equations are   compatible with a wide class of Lagrangians,  starting with a  modified Einstein-Hilbert Lagrangian as the simplest  one, provided  minimal boundaries are added  to the bulk.  A   non-trivial   canonical structure  is derived,  suggesting a canonical  quantization of the brane-world geometry  relative to the extra  dimensions, where the   quantum states  are set in  correspondence with   high frequency gravitational waves. It is shown that in   the cases of  at  least  six dimensions,   there exists a    confined gauge field included in the embedding structure. The  size of  extra  dimensions   compatible with  the embedding is  calculated  and found to be different from the one derived  with  product topology.
\end{abstract}


\vskip1cm]

\section{ Introduction}
The brane-worlds program  proposes a solution of the hierarchy problem at the TeV scale,
 assuming that the usual matter and  standard gauge interactions  remain confined  to a four-dimensional space-time embedded in a  higher dimensional bulk, while the   extra dimensions are  probed  by  gravitons. The size of the extra dimensions  is of the order of  tenths of millimeter,  as derived  from  the  effective   Planck scale in  four dimensions and  a  fundamental scale  in the bulk  set  at the TeV \cite{Arkani,Randall}.  

Brane-worlds inherits  its name   and some basic ideas from  Horava-Witten's M-theory, where   the  standard  model of interactions  contained in the  $E_{8}\times E_{8}$ heterotic  string theory is  also confined to a  3-brane, but  gravitons propagate  in the   11-dimensional bulk  \cite{string}.   However,  the use of large extra dimensions and  confined   gauge interactions in  higher dimensional   models  has been  considered  earlier,  under distinct motivations   \cite{KK}.  Also, the idea of  a space-time  embedded and    evolving in a  higher dimensional space  has  been proposed  in  various related applications,  such as  the generation of   internal  symmetries, quantum gravity, alternative  Kaluza-Klein theories and cosmology \cite{embedding}.  Recent  problems  and  reviews can be found in  \cite{problems,reviews}. 

In this note  we   attempt to answer some  basic  questions   which remain  open,  due in part to the fact that   most of  the  recent developments  are  specific to   particular models.  For  example this  has given the wrong impression that the    brane-world  program  is  necessarily  a  five-dimensional  theory based on the $AdS_{5}$ bulk,  or  that it  is  a higher  dimensional theory   defined on a bulk  with a product  topology. Thus  we   ask,  what is  the most general  geometrical scenario in which  such program  can  be  developed?  What are its essential, model independent,  postulates?   How  are  TeV gravitons  to be   defined  and  how  do    we  confine the  gauge interactions? Above all,  what is  the brane-world  action principle?

 As the program  stands today and   leaving aside   model dependent properties  such as  the $AdS_{5}$ bulk, warp factors  and  junction conditions,  we  may identify four  basic principles. One of   phenomenological  nature  sets the   fundamental  scale of  interactions   at  the TeV. The other  three  are  of  theoretical nature,  asserting that  the  extra dimensions are  probed by   TeV  gravitons, that  the  standard  gauge interactions  remain confined to the  four  dimensional space-time  and that  this  space-time is  embedded in  a higher dimensional bulk space.

Our purpose is to    avoid  the limitations  imposed by   the hypersurface  condition or by  the use of   specific  topologies,  studying   the  compatibility  between those principles in the most general situation,  assuming   that  the  bulk  has an arbitrary  number of  dimensions,  arbitrary signature and topology.  Thus,  brane-worlds  are  considered here in the   broader sense,   characterized only by  the  above basic principles. That is,  as  dynamically embedded  submanifolds,   such that they  retain   the   gauge  interactions confined within and that they  exhibit  some sort of quantum fluctuations. 

One of  our  results shows that   under  those very  general conditions  the Einstein-Hilbert  action arises naturally as the simplest  action derived   from the embedding  equations. We will see that  the total  divergence term  can be removed   before the application   of the variational principle, resulting in a considerable  simplification of the dynamics.

Although  very  little  has been said  about the  symmetries of the extra dimensions, we have not  found   any  motivation to  mix this   group with the   space-time diffeomorphisms.  Assuming that  these are   separate  symmetries,  we  are  able to 
derive  a canonical  formulation of  brane-worlds  and  sketch  a  model of  quantum theory.   

Another   result  shows  that when  the number of extra dimensions is greater than one and  that they admit  an isometry group, the  embedding equations  contain  a  confined  gauge-like potential,   whose gauge group is defined by that   symmetry. 

Finally,  we find that  the   size  of the extra dimensions  that  can be probed by gravitons,   compatible with   the embedding,    requires  the  existence of  minimal
boundaries. Using these boundaries we  find that  the size  differs  slightly  from that  estimated in  \cite{Arkani}.  However,   for  small incursions in a region where the embedding is smooth the  difference  is   negligible.  

 The paper is  organized as follows:
In section II brane-worlds are  described   from the  point of  view of geometric perturbations,  where each  perturbation  remains  an embedded  submanifold.    The   Lagrangian  for the higher dimensional space geometry is  derived   from  Gauss'  equation,  without  appeal to  any  particular  symmetry in  section III.  A  non-constrained canonical  structure  is  also derived. In section IV   we discuss the corresponding  quantum description of  a  brane-world   and the induced topological changes.  Section  V  shows  the confined gauge  field included in the embedding and its  implications to the   number of  the extra dimensions. The   size of these dimensions  compatible  with the embedding is  discussed at the end.   
 
\section{Geometric Perturbations and  Stability}

The  electromagnetic, weak and strong  interactions together with  the confined matter produce  tensions, pressures and energy in the brane-world,  which in turn  cause   deformations  on its   geometry.  Therefore,  a natural approach to  brane-world perturbative analysis  is to start  with  the perturbations of the   source, and   then find  the consequent  perturbation of the geometry.  To do this,  it  is  common practice   to  rely  on   junction conditions  relating the   energy-momentum tensor of the source  to the   extrinsic curvature.  Besides being  not unique,  it has been noted that   these conditions  are  difficult to solve  together with the other equations  \cite{junction}.  A    more general  procedure is  to  follow on the opposite direction, starting with the  perturbations of the geometry and  if  desired,  latter on   find the  perturbations of the   confined source \cite{Deruelle}.  This has the  advantage   of   requiring  a simpler dynamics
and  it can be applied to  any number of dimensions.  Actually,   by use of  Nash's perturbative  embedding procedure, we shall  see  that  the brane-worlds  may
 be  described  as  a family of stable  perturbations  of  a  given locally embedded  background  space-time. 

  The local embedding    is    constructed  in a neighborhood  of  each point  of the brane-world,    defining an embedding bundle whose total space consists  of    all  embedding spaces. Then, the    embedding  equations  are derived from the curvature tensor  of  each  local embedding space, written in the Gaussian frame  defined by  the embedded  submanifold and the normal vectors \cite{Eisenhart}.  From the point of view of brane-worlds, this  amounts to  have a   dynamic bulk  whose  geometry  depends on that of the brane-world,  as opposed to static or  rigid embedding.

Perturbations of   embedded submanifolds   with respect to  a transverse direction has been  used    as  a  way to generate  embedding theorems  along the following lines \cite{perturbations}:
Consider background $\bar{V}_{n}$  with metric  $\bar{g}_{ij}$, isometrically  embedded  in  $V_{D}$,   by a map $\bar{\cal X}: \bar{V}_{n}\rightarrow V_{D}$ such that\footnote{  
All Greek indices  run from 1 to $D$. Small case Latin indices  run from  1  to  $n$  and  capital Latin indices run from  $n+1$ to  $D$. The covariant derivative with respect to
the metric of  the higher dimensional manifold is denoted by  a semicolon and
$\xi^{\mu}_{;i}=\xi^{\mu}_{;\gamma}\bar{\cal X}^{\gamma}_{,i}$ 
denotes  its projection over  $V_{n}$.  The  curvatures of   $V_{D}$ are distinguished from that of  $V_{n}$ by  a calligraphic  ${\cal R} $. Since we have not fixed  the signature of $V_{D}$  the notation  ${\cal G}= |det({\cal G}_{\alpha\beta})|$ is  used throughout. }. 
\begin{equation}
\bar{\cal X}^{\mu}_{,i}\bar{\cal X}^{\nu}_{,j}{\cal G}_{\mu\nu} =\bar{g}_{ij},\;
\bar{\cal X}^{\mu}_{,i}\bar{\eta}^{\nu}_{A}{\cal G}_{\mu\nu}=0,\; 
\bar{\eta}^{\mu}_{A}\bar{\eta}^{\nu}_{B}{\cal G}_{\mu\nu}=g_{AB}   \label{eq:X}
\end{equation}
where we have denoted  by  ${\cal G}_{\mu\nu}$ the  metric of  $V_{D}$ in arbitrary coordinates  and  $g_{AB}$ denotes the components of the metric of  the complementary space orthogonal to $\bar{V}_{n}$, in the basis $\{\eta_{A}\}$. 
 The perturbations  of   $\bar{V}_{n}$  with respect to a small parameter $s$ along an  arbitrary transverse direction  $\zeta $  is  given by 
 \begin{equation}
{\cal Z}^{\mu}(x^{i},s) =  \bar{{\cal X}}^{\mu}+s\pounds_{\zeta}\bar{{\cal X}}^{\mu}
= \bar{\cal X}^{\mu} + s (  \zeta ,  {\cal  X}      )^{\mu}      \label{eq:deformation}
\end{equation}
The presence of   components of  $\zeta$  tangent to $V_{n}$  is  a cause for concern because   it can induce undesirable coordinate gauges. In geometric perturbations it is  possible to obtain coordinate gauge independency  simply   by selecting the   $\zeta^{\mu}$ to be orthogonal  to the background.  In this case, we obtain the  perturbations  of the  embedding map  along  a  single  orthogonal  extra  direction $\bar{\eta}_{A}$  as
\begin{equation}
{\cal Z}^{\mu}_{,i}(x,s^{A}) =\bar{{\cal X}}^{\mu}_{,i}(x)+ s^{A}\bar{\eta}^{\mu}_{A,i}(x).
\label{eq:Zi}
\end{equation}
Since  the   vectors  $\bar{\eta}_{A}$  are independent  and they  depend only of $x^{i}$,  it also follows that 
\begin{eqnarray}
\eta^{\mu}_{A}(x^{i})  =  \bar{\eta}_{A}^{\mu} +s^{B}[\bar{\eta}_{B},\bar{\eta}_{A}]^{\mu}=\bar{\eta}^{\mu}_{A} \label{eq:eta}
\end{eqnarray}  
However, it is not  obvious that   this  perturbation  represents a new  submanifold  or even  that  it is embedded in the same   $V_{D}$. For  example,   the  Schwarzschild space-time is known to be isometrically  embedded in  a  six  dimensional flat space with metric signature  $(4,2)$. Its  maximal analytic extension, the  Kruskal  space-time is  also embedded in  a six dimensional space, but with metric signature $(5,1)$ \cite{Fronsdal}. 
Now, the   Kruskal  space-time  may be seen as   a  perturbation of the  Schwarzschild  space-time  such that  it  becomes geodesically complete. Although the latter is  a subset of the former, they do not fit into the same flat bulk, unless the signature of the  six  dimensional space is  allowed to change. Therefore,  in the   general case  the geometry and topology of the bulk  should  not be fixed.

The integrability conditions  for the  perturbed geometry  are the  Gauss, Codazzi and Ricci equations, respectively 
\begin{eqnarray}
R_{ijkl}& =& 2g^{MN}k_{i[kM}k_{l]jN} +{\cal R}_{\mu\nu\rho\sigma}{\cal Z}^{\mu}_{,i}
{\cal Z}^{\nu}_{,j}{\cal Z}^{\rho}_{,k}{\cal Z}^{\sigma}_{,l}\nonumber \\
k_{i[jA;k]}& =& g^{MN}A_{[kMA}k_{j]iN} +{\cal R}_{\mu\nu\rho\sigma}
{\cal Z}^{\mu}_{,i} \eta^{\nu}_{A}{\cal Z}^{\rho}_{,j}{\cal Z}^{\sigma}_{,k}\label{eq:GCR}\\
2A_{[jAB;k]} &=& - 2g^{MN}A_{[jMA}A_{k]NB} \nonumber \\
 &-& g^{mn}k_{[jmA}k_{k]nB} - {\cal R}_{\mu\nu\rho\sigma}{\cal
Z}^{\rho}_{,j} {\cal Z}^{\sigma}_{,k}\eta^{\nu}_{A}\eta^{\mu}_{B} \nonumber
\end{eqnarray}
The  first two equations  have been   extensively applied to  the analysis of  brane-worlds in  five dimensions \cite{junction}, but as a  whole they have  not been appreciated in the case  $D\ge 6$. Assuming that  \rf{GCR} hold  true for  all perturbations,
the result is  an   $N$-parameter  family of  embedded  submanifolds  characterized by the  parameters  $s^{A}$, suitable for   a perturbative description of the brane-worlds, after implementing the  confinement  and the  quantization. 

The perturbation  \rf{Zi} and  \rf{eta}  induce  a  perturbation of the   metric $g_{ij}$   along  those  dimensions   which  can be written in general coordinates  as 
\[
g_{ij}=\bar{g}_{ij} +\chi_{ij}(x^{i},s^{A})
\]
In particular,  the linear  perturbation obtained from the expansion in   $s^{A}$  are
\[
g_{ij} =\bar{g}_{ij} +\epsilon^{A}\gamma_{ijA}(x^{i})
\]
where $ \epsilon^{A}$  is  a  small expansion parameter. Applying  this to Einstein's  equations  under   the de Donder gauge,    we obtain the linear    wave equation relative  to  the extra dimensions,  where the back reaction of  the background geometry  must be  taken into consideration. The  wave equation is  written in  the most general form as 
\begin{equation}
\Box^{ij}{}_{k\ell}\Psi_{ijA}(x,s) = 8\pi G { T}_{k\ell A} \label{eq:WAVE}
\end{equation}
where  $\Psi_{ijA} =\gamma_{ijA}- 1/2\gamma_{A} \bar{g}_{ij}$,  $\gamma_{A}= \bar{g}^{mn}\gamma_{mn}$  and  where (denoting$\bar{\nabla}_{k}\xi =\xi_{;k}$ for  clarity) 
\begin{equation}
\Box^{ij}{}_{k\ell}=\bar{g}^{ij}\bar{\nabla}_{k}\bar{\nabla}_{\ell} +2\bar{R}^{i\;\; \;j}_{\, k\ell} + 2\bar{R}^{i\;}_{ (k}\delta ^{\; j}_{\ell)}   \label{eq:Rahm}
\end{equation}
is the  generalized   (de Rahm)   wave operator,   containing  curvature terms  of the 
  background geometry.
  
Assuming that  the wave solutions of   \rf{WAVE}  correspond to   the  quantum modes of the brane-world geometry, they must  represent  gravitational  waves  of  high frequency. That is, with  a small wavelength  $\lambda$ as  compared with  a  local invariant  characteristic  length $\ell$ of the brane-world geometry, the curvature radius,  which  plays a  relevant role on the determination of the  classical modes.  This  radius  has been characterized as $inf |R_{ijkl}|$ \cite{hfv}, but in brane-worlds it   must be expressed  in terms of a  distance in the extra dimensions.  To find this  we  follow the  same  definitions as in  the geometry of  surfaces. Consider the  embedding equations of  the perturbed geometry  written in the  particular Gaussian  frame  defined by the  embedded  geometry and  the  $\eta^{A}$'s
 \begin{equation}
{\cal Z}^{\mu}_{,i}{\cal Z}^{\nu}_{,j}{\cal G}_{\mu\nu} =g_{ij},\;
{\cal Z}^{\mu}_{,i}\eta^{\nu}_{A}{\cal G}_{\mu\nu}=g_{iA},\; 
{\eta}^{\mu}_{A}{\eta}^{\nu}_{B}{\cal
G}_{\mu\nu}=g_{AB}   \label{eq:Z}
\end{equation}
where   $g_{iA}=s^{M}A_{iMA}$  and 
\begin{equation}
A_{iAB}  =\eta^{\mu}_{B;i}\eta^{\nu}_{A}{\cal
G}_{\mu\nu}= \bar{\eta}^{\mu}_{B;i}\bar{\eta}^{\nu}_{A}{\cal G}_{\mu\nu}
=\bar{A}_{iAB} \label{eq:AiAB}
\end{equation}
Replacing \rf{Zi} in  \rf{Z}, we may express the   perturbed metric in the Gaussian frame defined by the  embedding as
\begin{eqnarray}
g_{ij} = \bar{g}_{ij}  -2s^{A}\bar{k}_{ijA}
&+& s^{A}s^{B}(\bar{g}^{mn}\bar{k}_{imA}\bar{k}_{jnB} \nonumber\\
&+&  g^{MN} A_{iMA}A_{jNB}) \label{eq:g}
\end{eqnarray}
and  the  perturbed  extrinsic curvature  
\begin{eqnarray}
k_{ijA}  &=&       -{\cal Z}^{\mu}_{,i}{\eta}^{\nu}_{A;j}{\cal G}_{\mu\nu} \nonumber\\
&= & \bar{k}_{ijA} -
s^{B}(\bar{g}^{mn}\bar{k}_{miA}\bar{k}_{jnB} +g^{MN}{A}_{iMA}A_{jNB}) \label{eq:KijA}
\end{eqnarray}

The curvature radii  of the background $\bar{V}_{n}$ are   the  $n\times N$  values   $\ell^{A}_{i}$ of   $s^{A}$,  one for  each principal direction $dx^{i}$ and for each normal $\eta_{A}$,  satisfying the homogeneous  equation \cite{Eisenhart}
\begin{equation}
(\bar{g}_{ij} -s^{A}\bar{k}_{ijA})dx^{i}= 0  \label{eq:radii},  \;\;  A \mbox{ fixed}.
\end{equation}
The single curvature radius $\ell$ is  the smallest   of these  solutions, corresponding to the direction  in which the brane-world  deviates more  sharply from the  tangent plane. 
Considering all  contributions of  $\ell^{A}_{i}$, in such a way that  the smaller  solution of  \rf{radii} prevails,  the curvature radius  may be expressed as 
\begin{equation}
\frac{1}{\ell}= \sqrt{\bar{g}^{ij}g_{AB}\frac{1}{\ell^{i}_{A}}\frac{1}{\ell^{j}_{B}}}
\label{eq:rho}
\end{equation}
Since   \rf{g}  can also be written as   
\begin{eqnarray*}
g_{ij} =\bar{g}^{mn}(\bar{g}_{im}&- & s^{A}\bar{k}_{imA})(\bar{g}_{jn}-s^{B}\bar{k}_{jnB})\\  &+ & s^{A}s^{B}g^{MN}A_{iMA}A_{jNB}
\end{eqnarray*}
it follows that   the  components
\begin{equation}
\tilde{g}_{ij} =\bar{g}^{mn}(\bar{g}_{im}-s^{A}\bar{k}_{imA})(\bar{g}_{jn}-s^{B}\bar{k}_{jnB})  \label{eq:tildeg}
\end{equation}
 become singular  at the  solutions   of \rf{radii}.  Therefore, $g_{ij}$ and consequently, the metric  of the bulk  written in  matrix form  
\begin{equation}
{\cal G}_{\alpha\beta}=
\left(
\matrix{\tilde{g}_{ij} + s^{A}s^{B}g^{MN}A_{iMA}A_{jNB}   &   g_{iA}\cr
         g_{jB} &  g_{AB} } \label{eq:matrix}
\right)
\end{equation}
becomes also singular  at the points determined by those solutions. Of course, this  is  not   real singularity  of  $V_{D}$  but  a  property  of the Gaussian system defined by the brane-world $V_{n}$. However, this  singularity  breaks the continuity and regularity of  the integrability equations \rf{GCR}   which are  constructed  with  this system.  Therefore,  it  represents also a  singularity  for   the wave equation \rf{Rahm} which  depends on the  background geometry.  In short,  the  curvature radius $\ell$ sets a  local  limit  for  the  region in the bulk  accessed by  the   gravitons   associated with those high frequency  waves.

\section{Field Equations  for  Brane-Worlds}

Among the  three  independent   variables   $g_{ij}$, $k_{ijA}$ and  $A_{iAB}$  in \rf{GCR},  only   $g_{ij}$ is  normally  assumed  to propagate  along the extra dimensions. However, comparing  \rf{KijA} with the derivative of  \rf{g} we obtain   
the  generalized York's relation
\begin{equation}
\frac{\partial g_{ij}}{\partial s^{A}}=-2k_{ijA} \label{eq:YORKG}
\end{equation}
which shows that  the extrinsic curvature also propagates in the bulk,   as  a consequence of the   metric propagation.   Finally, from  \rf{AiAB} it follows  that  the  third variable  $A_{iAB}$  does not propagate. 

Since   we are not using  any particular  metric ansatz, we  must  follow a general procedure to  determine  the  variational principle  compatible  with  \rf{GCR}.  For that purpose we note that 
\begin{equation}
g^{ij}Z^{\mu}_{,i}Z^{\nu}_{,j}= {\cal G}^{\mu\nu}
-g^{AB}\eta^{\mu}_{A}\eta^{\nu}_{B}  \label{eq:jacobian}
\end{equation}
Using  this,   the contractions of  the first equation  \rf{GCR}  gives the Ricci scalar of  the perturbed geometry
\begin{eqnarray}
R  &=& (K^{2} -h^{2})  +{\cal R}  -2g^{AB}{\cal R}_{\mu\nu} \eta^{\mu}_{A}\eta^{\nu}_{B} \label{eq:RICCI}\\
& - & g^{AB}g^{MN}{\cal R}_{\mu\nu\rho\sigma}\eta^{\mu}_{A}\eta^{\sigma}_{B}\eta^{\nu}_{M}
\eta^{\rho}_{N} \nonumber       
\end{eqnarray}
where  $K^{2}=k_{ijA}k^{ijA}$  corresponds to the Gaussian curvature,   $h_{A}=g^{ij}k_{ijA}$ is the mean  curvature  for  $\eta_{A}$  and  $h^{2} =g^{AB}h_{A}h_{B}$.
In the Gaussian frame of the embedding  we may set $g_{AB}=\epsilon_{A}\delta_{AB}$,
$\epsilon_{A}=\pm 1$, so that  the last term in   \rf{RICCI} vanishes and 
\[
g^{AB}{\cal R}_{\mu\nu}\eta^{\mu}_{A}\eta^{\nu}_{B}  =-g^{AB}\frac{\partial
h_{A}}{\partial s^{B}} +K^{2}
\]
Since this  is  a tensor equation,  it holds  in any frame  and   \rf{RICCI} reduces to
\[
 R ={\cal R} - (K^{2} + h^{2})  -2  h^{A}_{,A}
\]
where the  divergence  can be  discarded under a  volume integration on $s^{A}$, provided  the mean curvatures  $h_{A}$ vanish at  given  boundaries. This  is  automatically satisfied  when  we assume that these boundaries  are minimal  submanifolds. These   fixed boundaries  replace the  dynamical boundaries used in  \cite{Randall}. With this,  after  discarding  this divergence  we obtain the   Lagrangian for  the brane-world
geometry
\begin{equation}
{\cal L}(g)= R\sqrt{g} ={\cal R}\sqrt{{ g}}+ (K^{2} + h^{2}) \sqrt{{g}}\label{eq:EH}
\end{equation}
Consequently,   the  dynamics of the   gravitational field in brane-worlds  follows from the Einstein-Hilbert   dynamics  of the bulk,  modified   by  the presence of the extrinsic  curvature term.

We  may also  construct   other   scalar invariants  with   contractions of  various  curvature terms  and their powers  to obtain  higher derivative  Lagrangians, or  even an  infinite series leading to  the Nambu-Goto  action.  The  modified Einstein-Hilbert  Lagrangian \rf{EH} is just the simplest one that can be derived from  the embedding equations \rf{GCR}, without further assumptions.

The use of  a variational principle permits us to  introduce  an  independent source in the brane-world. Thus  we add  to  \rf{EH} the  Lagrangian of the  confined  matter   ${\cal L}_{m}$.  Then, the  field  equations  with respect to the  metric  $g_{ij}$, with the confined  matter represented  by  $T_{ij}^{m}$ are
\begin{equation}
R_{ij}\!\! -\!\frac{1}{2}R g_{ij}\!  = \! 8\pi GT^{m}_{ij}\! +\! ({\cal  R}_{\mu\nu}\!-\!\frac{1}{2}{\cal R}{\cal  G}_{\mu\nu})Z^{\mu}_{,i}Z^{\nu}_{,j}\!  +\! Q_{ij} +  S_{ij} \label{eq:Einstein}
\end{equation}
where  we have denoted
\begin{equation}
Q_{ij} =g^{AB}(b^{m}_{iA}b_{j m B} -h_{A}b_{ij B})  -\frac{1}{2}(K^{2}- h^{2})g_{ij}
\end{equation}
and  
\begin{equation}
S_{ij}  = g^{AB}{\cal R}_{\mu\nu} \eta^{\mu}_{A}\eta^{\nu}_{B}g_{ij} - \frac{1}{2}g^{AB}{\cal R}_{\mu\nu\rho\sigma}\eta^{\mu}_{A}\eta^{\sigma}_{B}Z^{\nu}_{,i}Z^{\rho}_{,j}
\end{equation}
The  value of  $ ({\cal  R}_{\mu\nu}-\frac{1}{2}{\cal R}{\cal  }_{\mu\nu})$   depends on  the definition of the  geometry of the bulk, generally  taken as  a  solution of the higher dimensional Einstein's  equations,  with a  bulk source represented by the  energy momentum tensor   $T_{\mu\nu}^{B}$.  In this case, due to  presence of the  factor   $Z^{\mu}_{,i}Z^{\nu}_{,j}$,  at the end    this bulk  matter is  projected onto  the  brane-world  in accordance with the  confinement hypothesis.  

In some  models  with a particularly chosen bulk geometry, the last term    $S_{ij}$  vanishes. On the other hand the   term  $Q_{ij}$   depends essentially  on the extrinsic curvature  and it does not  necessarily vanish, even  if the bulk is flat. Therefore,  this term  may effectively   modify  the usual  Einstein  dynamics.  We will discuss  its meaning  in a cosmological application in a subsequent paper.

The   solutions of  \rf{Einstein}  describe   the gravitational  field in the   brane-world,   showing  the   additional   terms  resulting from the embedding.   This must be  complemented   by  the  description of the evolution of the  brane-world geometry  along the extra  variables. For this  purpose,   we need  a   canonical  structure    compatible with  \rf{Einstein}  and   compatible with the  perturbative     analysis  of  section II.

Unlike   general relativity,   the  coordinates on the extra  dimensions    behave  as  the propagation parameters for the embedded geometry, so that  the   phase space   has to be    defined   with respect  to these  parameters. Thus, the momentum  conjugated to ${\cal G}_{\alpha\beta}$, relative to the extra dimension   $\eta_{A}$ is  given by 
\[
p^{\alpha\beta}{}_{ (A)} =\frac{\partial {\cal L}}{     \partial \left(
\frac{ \partial {\cal G}_{\alpha\beta} }{\partial s^{A}}   \right)     }
\]
and  in particular, using  \rf{YORKG} we obtain the components
\begin{equation}
p^{ij}{}_{(A)}=-(k^{ij}{}_{A} +h_{A}g^{ij})\sqrt{{\cal G}} \label{eq:MOMENTUM}
\end{equation}
which  corresponds  to the  propagation of   $g^{ij}$ along  $\eta^{A}$.

The  confinement  hypothesis implies that   any gauge fields and  matter sources  which could  eventually be contained in  the  extra  components of the  bulk metric 
should not   propagate.  Therefore, consistently with this  we add the momentum constraints
\begin{eqnarray}
p^{iA}{}_{(B)}  &= &  -2\frac{\partial{\cal
R}_{\alpha\beta}\eta^{\alpha}_{A}\eta^{\beta}_{B} }{\partial \frac{\partial{\cal
G}_{iA}}{\partial s^{B} }} \sqrt{{\cal G}}=0  
\label{eq:DEFCON1},\\
p^{AB}{}_{(C)} & = & -2
\frac{\partial {\cal R}_{\alpha\beta}\eta^{\alpha}_{A}\eta^{\beta}_{B}}
{\partial \frac{\partial {\cal G}_{AB}}{\partial s^{C}}}\sqrt{{\cal G}}=0 \label{eq:DEFCON2}
\end{eqnarray}
These constraints   are also   consistent with our previous choice of   orthogonal  perturbations
 given by \rf{Zi} and  \rf{eta}.   

Using \rf{DEFCON1} and \rf{DEFCON2}, the Hamiltonian  corresponding to  the displacement  along a  single  direction  $\eta^{A}$,  follows from the  Legendre transformation 
\begin{eqnarray}
{\cal H}_{A}(g,p) &=&   p^{ij}{}_{(A)}g_{ij,A}-{\cal L} =\nonumber \\
 &-& R\sqrt{{\cal G}}  -\frac{1}{{\cal G}}\left(\frac{p_{A}^{2}}{n+1}
- p_{ij (A)}p^{ij}_{(A)} \right) \label{eq:HA}
\end{eqnarray}
where  we have denoted  $p_{A}=g_{ij}p^{ij}{}_{(A)}$.    Hamilton's equations  relative to the extra  coordinate  $s^{A}$ are
\begin{eqnarray}
& &\frac{d g_{ij}}{d s^{A}}  = \frac{\delta {\cal H}_{A}}{\delta p^{ij (A)}}=
\frac{-2}{\sqrt{\cal G}} \left(\frac{g_{ij}p_{A}}{n+1}-p_{ij(A)}\right),\vspace{3mm}\label{eq:DOTG}\\
& &\frac{d  p^{ij}_{(A)} }{ d s^{A} }  = -\frac{\delta {\cal H}_{A}}{\delta g_{ij}}= (R^{ij}-\frac{1}{2}Rg^{ij})\sqrt{\cal G} +\nonumber\\
 & & \frac{1}{\sqrt{\cal G}}\left[ \frac{2p_{A}p^{ij}{}_{(A)}}{n+1} \right. 
+ \left. \frac{1}{2}\left(
\frac{p_{A}^{2}}{n+1} +p_{mn(A)}p^{mn}{}_{(A)} \right) g^{ij}\right] \label{eq:DOTP}
\end{eqnarray}
The first of these equations is  the same as   York's  relation \rf{YORKG} expressed in terms of  $p^{ij}{}_{(A)}$,   giving the propagation of the metric in terms of  the extrinsic curvature.  The second equation   expresses  the propagation of the  extrinsic  curvature  expressed in terms of  $p^{ij}{}_{(A)}$.

We  conclude  that  the Hamiltonian  dynamics  expressed by  \rf{HA} through \rf{DOTP}  describe the same a  motion which  equivalent  to  the one given by the   perturbative analysis  in  section II.   

\section{Quantum States} 

The compactification of  the extra dimensions  down to  Planck's length was introduced to make  Kaluza-Klein  theory  compatible with  quantum mechanics, where the normal modes  of the harmonic  expansion  with respect to the  internal parameters  were set in correspondence with  quantum modes \cite{Klein}.  As we know, in that theory   the   strong curvature of the internal space  contributes to large  mass  fermion  states, which are not observed at  the electroweak scale.  If     the extra dimensions    were  large or   non compact,
then we   would obtain massless or   light Kaluza-Klein  modes, which could be observed at  that energy  scale.  However,   it is  not  clear  that these modes  would still   keep a  correspondence  with  quantum states. 

Contrasting with the Kaluza-Klein program,  in  brane-worlds  only  the  gravitational field is   expanded  along the  extra dimensions,  with   modes associated with gravitational  waves \cite{Chamblin}. Then the  fermion chirality problem would not arise  but  the  metric  expansion  should hold independently of the fact that these  dimensions are large,
compact or  not. In other words,  the   quantum correspondence    must  be independent  of the  bulk topology but it must be compatible  with the embedding. As remarked before,  the gravitational  waves  associated with the   quantum fluctuations of the geometry make sense only  in the high frequency limit,  which  depend   on the local  geometry  of the background,  and not   on  the topology of the bulk.  

In the previous  sections  we  have seen  that  the same perturbations that lead to the  wave equation   also  lead  to  a    canonical formulation  derived from the Hamiltonians  \rf{HA}. Consequently,  the quantum states   associated with   the  high  frequency waves   can be,  at least in principle,   defined  by the    canonical quantization   defined by 
those Hamiltonians  with respect to the extra dimensions.   The  procedure  would  be  similar to that of  the ADM  formulation of general relativity,  with an important difference:  Since the   extra  dimensions do not transform  under  the same diffeomorphism group   of the brane-world,  the  Poisson bracket structure   does not suffer the  same propagation problem. Instead, it  behaves differently  under the   brane-world  diffeomorphisms and  under the  transformations of the extra coordinates. 
Therefore, the   evolution of  a  functional  ${\cal  F}$ in phase space  relative to a  single extra dimension  $\eta_{A}$,  given in terms of  Poisson brackets  as
\[
[{\cal F},{\cal H}_{A}]=\frac{\delta {\cal F}}{\delta g_{ij}}\frac{\delta {\cal H_{A}}}{\delta  p^{ij(A)}}-\frac{\delta {\cal F}}{\delta p^{ij(A)}}\frac{\delta {\cal H_{A}}}{\delta  g_{ij}}=\frac{\delta  {\cal F}}{\delta s^{A}},
\]
propagates  covariantly  along  the  evolution of the system. Thus, a  canonical quantization may be defined for  each separate  ${\cal H}_{A}$   associated with an operator  $\hat{\cal H}_{A}$  acting on a Hilbert space, where the  quantum state  of the  embedded brane-world is  given by the wave function  $\Psi_{ij(A)}$  and the  final  state  is  given by a  superposition  $\Psi_{ij}=\sum_{A} \alpha^{A}\Psi_{ij(A)}$.

The  wave functions    $\Psi_{ij(A)}$    describe    spin-2  fields  in   the  brane-world as  solution of the  Klein-Gordon-like  equation  associated with the  de Rahm  operator  \rf{Rahm}.  However, its   evolution    along the  extra  dimensions   requires the  explicit   variation of these functions with  respect to  $s^{A}$.  Since they  do not mix  with the   brane-world  coordinates,  they  may  be used  as  time parameters.  As  a   naive example consider that  the   quantum states  are described  by  Sch\"odinger's  equations   with respect to   $s^{A}$
\begin{equation}
-i\hbar \frac{ d \Psi_{ij(A)} }{ d s^{A}} =\hat{\cal H}_{A}\Psi_{ij(A)}
\label{eq:Schroedinger}
\end{equation}
Then the  probability  of  a  brane-world to be  in an  embedding  state  $\Psi_{ij(A)}$  is   given by 
\[
||\Psi_{ij(A)}||^{2}=\int \Psi_{ij(A)}^{\dagger}\Psi_{ij(A)}\,dv
\]
where  the integral extends over a volume  in $V_{D}$  with a base on a  compact region 
of  $V_{n}$ and  a finite extension of the extra coordinates, such that  it does not break the limit   $\ell$ of  regularity of the embedding functions.

Topological changes such as  the emergence of  handlers, black holes  and wormholes, induced by  the probability transitions,  are expected to occur  from  high energy oscillations \cite{top}.  Thus, for  example, if   $\eta_{A}$ and  $\eta_{B}$ are  both space-like  extra  dimensions, then,  the  classical limit of the probability transition   $<\Psi_{ij\, (A)},\Psi_{k\ell\, (B)}>$   corresponds   to a transition from a perturbation  of  $\bar{V}_{n}$ along  $\eta_{A}$  to  a  perturbation along  $\eta_{B}$.   An observer  in  $V_{n}$  may  interpret the result  as the emergence of  a space-like  handle.  On the other hand, if  $\eta_{A}$ and  $\eta_{B}$  have  both time-like signatures, then the classical limit  would    correspond to a closed loop involving  two internal time-like parameters. 
 
When   $\eta_{A}$ and  $\eta_{B}$  have different signatures,   the transition probability  must  also  take into account  possible changes of  signature.  Considering  again the   Kruskal brane-world  example,   regarded as  a geodesically complete  perturbation of the Schwarszchild  space-time,  we may fit both space-times  in the same dynamical  six dimensional flat space, provided  a  quantum signature transition at the  horizon is  considered.

\section{Confinement  of gauge  interactions}

Since most  of current  discussion on brane-worlds is concentrated on  models with just one  extra dimension, not much  has been said  about  the  symmetries of the   extra  dimensions.
In  strings  or M-theory  all internal symmetries  derive from the  string group (e.g. $E_{8}\times E_{8}$ or  $SO(32)$)  so that  additional symmetries on the extra dimensions are not required or  even  wanted. Quite on the opposite direction,  Kaluza-Klein theory   with  a ground  state like  $M_{4}\times B_{N}$ requires a  maximal symmetry for the space  $B_{N}$ generated  by the extra dimensions.  In brane-worlds  the gauge  interactions remain confined  independently of the   state of the  embedded geometry, suggesting  that  the gauge group should  also be  independent of the  embedding  state.   

The  bending  of  brane-worlds is  described by the  variation of the normal vector  $\eta_{A}$  when  its foot  is displaced along the  brane-world.    In general   it   has   tangent and  normal  components   with coefficients  $k_{ijA}$ and  $A_{iAB}$  respectively. The variation of  $k_{ijA}$ produces  a tension  on the brane-world and consequently a change  of the energy-momentum tensor  of the confined source \cite{junction}. On the other hand, as  evidenced by \rf{AiAB}
$A_{iAB}$  does not  propagate  with  $s^{A}$. In  order to understand  its meaning, consider  the cases  $D\ge 6$  and that  the  space  generated by the extra dimensions has a  certain number of   Killing vector fields. Then  we may apply the  
relevant, but  little explored fact that   $A_{iAB}$ transform  as the components of a   gauge potential  under  that  group  of  isometries. This  can be  seen  from  the  transformation of the  mixed component of the  metric tensor, of  $V_{D}$
under a local  infinitesimal   coordinate  transformation of  the extra coordinates  but leaving 
fixed the coordinates of  $V_{n}$: 
\[
 s'^{A}= s^{A} +\xi^{A}\;\;\mbox{with}\;\;  \xi^{i}=0,\;\; \mbox{and}\;\;  \xi^{A}  =\theta^{A}_{M}(x^{i})s^{M}
\]
where  $\theta^{A}_{B}$  are  infinitesimal parameters.
Denoting   generic coordinates  in  $V_{D}$ by $\{x^{\mu}\} =\{x^{i},s^{A}\}$, it follows that  \cite{Maia}
\[
g'_{iA} =g_{iA} + g_{i\mu}\xi^{\mu}_{,A} +g_{A\mu}\xi^{\mu}_{,i} +\xi^{\mu}\frac{\partial  g_{iA}}{\partial x^{\mu}} +0(\xi^{2})
\]
Therefore the  transformation of  $A_{iAB}$  follows from
\[
A'_{iAB}=\frac{\partial  g'_{iA}}{\partial  s'^{B}}= \frac{\partial  g'_{iA}}{\partial  s^{B}}
- \xi^{\mu}_{,B}\frac{\partial g'_{iA}}{\partial x^{\mu}}
\]
Using  $\xi^{A}_{,B}=\theta^{A}_{B}(x^{i})$ and  $\xi^{A}_{,i} =\theta^{A}_{B}s^{B}$ we obtain
\begin{equation}
A'_{iAB} =  A_{iAB} -2g^{MN}A_{iM[A}\theta_{B]N} +g_{MB} \theta^{M}_{A,i}
\end{equation}
showing  that in fact  $A_{iAB}$ transform  as  the components of  a non-Abelian gauge potential, where the gauge group is the  group of  isometries of the   extra dimensions.  This property   strongly suggests  that   $A_{iAB}$   should  be   considered  as    a  confined gauge potential  when  $D\ge 6$ and the  extra  dimensions have an isometry group.  

The simplest  embedding theorem   concerns the  analytic  embeddings in  flat spaces  $M_{D}$ \cite{Cartan}. The analytic assumption greatly simplifies the embedding   and it  implies that $10$ dimensions are sufficient and  it can be done in  even less dimensions. 
However, in brane-worlds the oscillations  of the embedded  geometry  are  taken as solutions of the  differential equations \rf{GCR},   and the   analyticity  implies  that these oscillations  are  represented by  convergent positive  power series. This  represents  a   limitation of   the  spectrum of  solutions,  including  the  probing  near  singularities where the power  series  may become divergent.  To avoid  these limitations   we  assume that these oscillations correspond to   differentiable  solutions of   \rf{GCR}.  In this case,  a more  powerful  embedding  theorem shows that  the limiting  dimension for  a flat embeddings  rises  to $14$,  or, more generally   for   an n-dimensional  submanifold   $  n(n+3)/2$,  with a   wide range of compatible signatures \cite{Greene}.
 
Consequently,  with the exception of five-dimensional bulks,   we  may   use  the   gauge  degree of  freedom to determine the  number of  extra dimensions,  such that  $A_{iAB}$  is  the  confined  gauge potential.  
Taking the  standard model   $SU(3)\times SU(2)\times U(1)$ acting on a seven-dimensional projective space, identified with  the space  generated by the extra  dimensions, we obtain  as in  Kaluza-Klein and  supergravity theories an  11  dimensional  space, which  may be  realized  in a flat bulk. On the other hand,  it  has  been suggested that  the new physics occurring at the TeV may  require  a larger  gauge group \cite{BH}.  If we take this into account  along with the motivations for  $SO(10)$   GUT,  the differentiable 
 embedding   gives  a   fourteen-dimensional  flat  bulk   with signature $(11,3)$  where  $A_{iAB}$ acts as a  self-contained  and confined  $SO(10)$  gauge field.  

Regardless of the topology of the extra dimensions we  need to  know   how  far  these dimensions can be   probed  by gravitons.  Currently there  are two approaches to this problem:  In \cite{Randall},  the volume of the space probed by  gravitons is   determined  by   the addition  of  two  boundary terms to the Einstein-Hilbert Lagrangian.    A  radion field  included in the   metric  takes care of the    separation between the boundaries.  On the other hand, the derivation of  the  size  in \cite{Arkani}  assumes that the  bulk has a fixed product topology,  where   the  volume of the extra dimensions  is  finite.  The  case  of  a single extra dimension is  excluded  because it   leads to a very  large extra dimension.

   Here, for generality  we have not  specified  a  metric  ansatz  and   for  compatibility with the embedding we  have  not  imposed  any  topological condition on the bulk.  To  find  the size  of  the extra dimensions under these  general  conditions,  take  a compact region  in $\bar{V}_{4}$  and  a  finite volume $\cal V$  of  the space  generated by the extra dimensions  limited by  two minimal  boundaries  for the variables $s^{A}$,  such that this region is  effectively probed by gravitons. From our previous discussion,  to keep  the  regularity of embedding and  wave  equations  we   require that  the length of the extra dimensions should not exceed  $\ell$.  Thus, the action integral for the  brane-world in this region using  \rf{EH} is (for  $n=4$)
\[
 \int\!\!\!\!\int R \sqrt{g} d^{4}x d^{N}s = \! \int {\cal R}\sqrt{g} d^{D}x -\int\!\!\!\!\int (K^{2}+ h^{2})\sqrt{g} d^{4}x d^{N}s
\]
where we notice that   all integrands   depend  on $x^{i}$ and $s^{A}$,   so that the indicated integrals  cannot be separated.  However, for  small oscillations of the brane-world such that  $ (s^{A})^{2}<<s^{A}<\ell $,  and  using  \rf{matrix} in an appropriate frame   we obtain  ${\cal G}= g$, so that the Einstein-Hilbert action for the bulk  is
\[
\int\!\!{\cal R}\sqrt{\cal G}d^{D}x\approx   \int\!\!\!\!\int\! R \sqrt{g} d^{4}x d^{N}s +\!\!\int\!\!\!\!\int (K^{2}\!+\! h^{2})\sqrt{g} d^{4}x d^{N}s
\]
However, from \rf{YORKG}  we see  that    $g_{ij}$ still has  a linear  dependence  on  $s^{A}$.   This can be  eliminated  without further  impositions on  the bulk   by selecting  a  sufficiently smooth  background  at the embedding  neighborhood, so that  $\bar{k}_{ijA} \approx 0$.  With this choice,  using  the  same  arguments  as in \cite{Arkani}   we may write
similarly to  \cite{Arkani}
\[ 
\frac{1}{M_{*}^{2+N}} \approx (\frac{1}{M_{Pl}^{2}} + \frac{1}{M_{e}^{2}}){\cal V}
\]
where   $M_{*}$  and  $M_{Pl}$  are respectively the fundamental and  effective scales, and where we have introduced   an   extrinsic  (or  bending)    energy scale  given by
\[ 
 \frac{1}{M_{e}^{2}}=\int (K^{2} + h^{2})\sqrt{g}d^{4}x
\]  
This is not necessarily  zero,  even when the   brane-world is flat.  As such it  may represent an observable  effect on the   brane-world  dynamics in the form of the  extrinsic tensor  $Q_{ij}$ in \rf{Einstein}. 

  Denoting  by  $d$  the typical length of the  extra dimensional space probed by the gravitons,  we may set
 ${\cal V}\approx d^{N}$  and under the specified conditions  we obtain
\[
d \approx   \frac{ M_{Pl}^{2/N}}{M_{*}^{1+2/N}}\frac{1}{(1 +  \frac{M_{Pl}^{2}}{M_{e}^{2}})^{1/N}}
 \]
The  size predicted  in \cite{Arkani} and  \cite{Randall} is recovered  when  $M_{Pl}^{2}<<M_{e}^{2}$  which occurs  with the  suggested  approximations. We cannot  make  such approximation  in general without  imposing  limitations  on the  brane-world  oscillations. 

When  we  have  several extra dimensions  the bending  is  determined by $k_{ijA}$ and  the gauge field $A_{iAB}$.  This    eliminates   the need to introduce a radion field in the brane-world  metric which  appear in  the hypersurface cases.

\section{Summary}

We  have investigated the most   general   geometrical scenarios in  which  a brane-world program compatible with the hypotheses of embedding, confinement and the existence of quantum states  can be implemented.  Our analysis is  independent of  any previous choice of  geometry, topology,  number of  dimensions and signature  for the  bulk.  Instead,  we have used the natural assumption that  the  brane-world   geometry must remain a local differentiable   embedded  submanifold  oscillating  between   minimal  boundaries.  We   have found that   the  four  basic postulates are sufficient to go  a long way towards  the  formulation of  a brane-world theory,  but some conclusions apply only  when the bulk  has  at least  six dimensions. 

Our  first result consists in the  derivation of  a general  dynamical principle   for  brane-worlds.  We have shown that  the  Lagrangian for the brane-world geometry  differ  from the Einstein-Hilbert Lagrangian  by  a term, which depends  essentially of the extrinsic curvature.  The implication of  this  is  that in general the bulk responds to the dynamics of the brane-world  and  consequently it  should be  allowed to have a variable  geometry.

It  is  also  possible  to add to  the modified Einstein-Hilbert basic  Lagrangian  \rf{EH} a  constant term and powers  of  scalar  functions  constructed with the  curvature tensors  derived from \rf{GCR},  to obtain  higher derivative  Lagrangians  provided we  also take in  account the corresponding  powers of the extrinsic curvature term.  

Using the fact that  the  extra dimensions do not obey  the same symmetry 
 of the four-dimensional  brane-world,   we  have managed to derive    a  non-trivial canonical structure and suggested   a  canonical  quantization of the  brane-world  relative to  the extra  dimensions based on the Hamiltonians  ${\cal H}_{A}$. 

When the  bulk has  at least  six  dimensions, a  confined  gauge field   is  contained in 
the   embedding  structure. This novel confinement mechanism   appear in the form of  one of the   basic embedding  variables  $A_{iAB}$. When we identify this field with the   physical   gauge field, a simple arithmetic  fixes  also  the number of extra dimensions:  For  the standard model it was found that  the  self-contained gauge structure  requires   11  dimensions. 
On the other hand,  the $SO(10)$  gauge group  implies in 14  dimensions,  which  can be  realized by a  flat bulk. 

Five dimensional, or more generally hypersurface  models are not excluded  from our  analysis but   since  they do  not  contain the field  $A_{iAB}$,   the confined  gauge fields need to be introduced  by other mechanisms. In this case, the equations  can be derived from the  general case  by setting $D=n+1$,  $A,B\cdots  =n+1$,  $g_{AB}=g_{n+1\,n+1}=\pm1$ and  $g_{i\, n+1}=0$.  Only the  first two   equations  in \rf{GCR} remain and are required  to  obtain a  Lagrangian similar to \rf{EH},   suitable  to  describe the   evolution of the brane-world with respect to the single   extra dimension.     

One  difficulty  associated with perturbations of hypersurface brane-worlds in a  constant curvature  bulk is  due to a   general result in geometry,  stating  that if  a hypersurface  has more than two finite curvature radii  $\ell_{i}$, then  it becomes indeformable \cite{Eisenhart}. This means that  there is  a certain degree of stiffness associated with perturbations,   preventing the  generation of  more complicated configurations of the embedded geometry. 

 The typical size of the extra dimensions   compatible with the embedding  was found to be close to the one predicted with product topology,  as long as   we remain in the  linear regime of  perturbations in a very smooth background.

We have not included  some relevant questions   such as the emergence of  a  cosmological constant  and  the observable  implications of the extrinsic terms in the dynamical  equations  \rf{Einstein}. Problems  related to  brane-world  cosmology  become  extremely interesting  under the  Lagrangian  \rf{EH}, where the extrinsic  geometry   contributes to the   modification of  Friedmann's  equation,  as   will be  discussed in a subsequent paper.

\acknowledgements{The authors  wish to acknowledge  the discussions on the subject with Drs. P.  Caldas  and V. Silveira.}

\end{document}